# Quantum Optimal Control and Level Sets

**Fariel Shafee**
Department of Physics
Princeton University
Princeton, NJ 08540

*Abstract:*

We investigate how the concepts of optimal control of measurables of a system with a time dependent Hamiltonian may be mixed with the level set technique to keep the desired entity invariant. We derive sets of equations for this purpose and also algorithms for numerical use. The notion of "constancy" of measurables in this context is also examined to make the techniques more useful in a real-life situation where some variability of the measurable may be tolerable.

## I. INTRODUCTION

Optimal control of quantum systems [1-4] involves the design of a control Hamiltonian with simultaneous satisfaction of a number of criteria, among which may be the difference of the expectation value of a target operator from a pre-assigned value, the minimal cost of energy required for control, and other relevant constraints, including of course, satisfaction of the schrodinger equation.

The usual technique involves beginning with a good guess of the control Hamiltonian in conformity with experimental practice and possibilities, e.g. with laser pulses we may use a frequency and amplitude of reasonable values. Then the cost function accounting for all the subcosts for the different constrains are calculated, and then variations using a genetic algorithm or some other method are tried to find an optimum. Schrodinger equation may be put in as a constraint in the cost function using a Lagrange multiplier, or it may be used independently to calculate trial wave functions.

The level set method [5-7] is a technique to study the propagation of curves or surfaces of constant values of a given quantity. Hence it seems reasonable to expect that it has relevance to the optimal control problem. We have previously [8] considered the time independent problem, where time was replaced by a scale parameter and the method was used in the restricted sense of reconstructing a surface from a finite number of data points. In this work we examine the more interesting situation where the parameters of the system Hamiltonian as well as the control field are all time dependent.

## II. QUANTUM LEVEL SET FORMALISM

Let the total Hamiltonian including both the system and the control part be given by

$$H = H_{sys} + H_c \quad \dots\dots\dots\dots\dots\dots\dots\dots\dots\dots\dots\dots\dots\dots\dots\dots\dots(1)$$

where the control part may be of the form

$$H_c = -\mathbf{\mu} \cdot \mathbf{E} \quad \dots\dots\dots\dots\dots\dots\dots\dots\dots\dots\dots\dots\dots\dots\dots\dots\dots(2)$$

with $\mathbf{\mu}$ the dipole moment and $\mathbf{E}$ the applied electric field of the laser. Let the system Hamiltonian be a function of $n$ number of time dependent parameters $a_i(t)$, which we write as the vector $\mathbf{a}$. So $\mathbf{a}(t)$ actually traces out a curve in $\mathbf{R}^n$ and $d\mathbf{a}(t)/dt$ is the velocity vector in parameter space which is tangential to the curve for each $t$. If we leave out $a_n$ to be the control parameter, the remaining components of $\mathbf{a}$ trace out a curve in $\mathbf{R}^{n-1}$.

Now if we consider any measurable $\Theta$ and a small time interval:

$$\langle t|\Theta|t\rangle = \langle t_0| e^{iH(t-t_0)} \Theta e^{-iH(t-t_0)}|t_0\rangle = \langle t_0|[H,\Theta]|t_0\rangle \, i\,(t-t_0) \quad \dots\dots\dots\dots\dots(3)$$

which gives

$$d\mathbf{a}(t)/dt \cdot \mathbf{grad}_a <\Theta> = id\mathbf{a}/dt \cdot <[\mathbf{grad}_a H, \Theta]> \quad \text{...............................................(4)}$$

For constant $<\Theta>$, with no explicit time dependence in the $<\Theta>$, the right hand side of Eq. 4 is zero, as the level set method finds. On the other hand it should normally be simple to find the operators

$$\mathbf{H_a} = \mathbf{grad}_a H = \mathbf{grad}_a V \quad \text{...........................................................................(5)}$$

and its commutator with $\Theta$. Say

$$\Theta_a = i[\mathbf{grad}_a V, \Theta] \quad \text{...................................................................................(6)}$$

is known and hence we should be able to get

$$d\mathbf{a}/dt \cdot <\Theta_a> = 0 \quad \text{.............................................................................................(7)}$$

which is the level set equation in the quantum context. We see that if $a_i$ [i = 1,…n-1] are the system parameters and we know the time dependence of these parameters, and $a_n$ is the control parameter [for example the field $\mathbf{E}(t)$], then in principle, Eq. 7 allows us to calculate the shift necessary in $a_n$ to move from t to t +dt, keeping $<\Theta>$ constant, i.e. we can integrate this to move from the level set at one t to the level set for another t, or equivalently we know how to get the time dependence of the field E(t) from the known quantities in the equation . Fig .1 shows the movement of the system from one $(a_1(t_1), a_2(t_1))$ point corresponding to a particular $a_3(t_1)=c_1$ giving the required $<\Theta>$ to another point $(a_1(t_2), a_2(t_2))$ corresponding to another value of the control parameter $a_3(t_2)=c_2$, but the same value of $<\Theta>$. In other words the time parameter is a reparameterization of the third parameter $a_3$, so that in the $(a_1, a_2)$ plane each t curve can also be labeled by a different $a_3$.

However, it is not sufficient for the operator $\Theta$ to be explicitly time independent for $<\Theta>$ to be explicitly time independent, because the expectation value carries time dependence through the time dependence of $|t>$.

$$\partial_t <\theta> = -i<[H,\theta]> = -i<\theta_o> \ (say) \quad \text{...........................................(8)}$$

Hence the complete Level Set equation is

$$\Theta_0 - u_a \cdot \Theta_a = 0 \quad \text{.............................................................................................(9)}$$

## III. OPTIMAL QUANTUM CONTROL

As we remarked earlier it is possible that there are other constraints besides keeping $\langle\Theta\rangle$ invariant. In that case we enter the domain of optimal control. The algorithm for OCT is fairly well laid out by now as we have pointed out in the Introduction. However, we should also remember that in usual OCT one postulates attaining the desired state after a finite time T, whereas in the previous section we have considered a continuous dynamics of the system with constant $\langle\Theta\rangle$. In reality data points can be obtained only over a finite mesh and hence any resolution exceeding the step size used is spurious except for considerations of smoothness, which is a desirable property in most physical models. If a laser beam is used for control, we have one time scale given by the frequency of the electromagnetic field and another by the duration of the pulse. In the level set method we have used virtually a smooth non-oscillating system.

Suppose we can tolerate $\langle\Theta\rangle$ given by

$$\langle\Theta\rangle = \langle\Theta\rangle_0 + \langle\Theta\rangle_1 \sin[\omega t] \quad\quad\quad\quad\quad\quad\quad\quad\quad\quad\quad\quad\quad\quad\quad\quad\quad\quad (10)$$

with the second term much smaller than the first, then we still have a nearly constant $\langle\Theta\rangle$, even more so if the time resolution we are interested in is coarser than $1/\omega$. In place of a simple sinusoidal periodic dependence we may have more complicated ones limiting $\langle\Theta\rangle$ to a given range around the desired value.

If we introduce three time scales, we can have
i) the frequency of the laser radiation giving $[t_0 \sim 1/\omega_{rad}]$
ii) the duration of the laser radiation $[n\, t_o$, where n is the number of wavelengths in the pulse], about an order of magnitude longer.
iii) the time averaging of observing $\langle\Theta\rangle$ : [ T ] ,

and then we can use the standard formalism of ref. [1] using Lagrange multipliers to minimize the difference of $\langle\Theta(T)\rangle$ from a target value, which would be equivalent to having a "constant" $\langle\Theta\rangle$ within our experimental resolution, and at the same time we can satisfy schrodinger equation, and the cost optimization constraint on the intensity of the laser radiation mentioned in ref. [1].

Fortunately, it appears that we can still use the level set method to directly relate the required field intensity (at a finer time scale as mentioned) to the variation of the system parameters with time) obviating the use of expensive inversion algorithms.

In case we want to study not only a few parameters of the potential [like effective depth and width and separation in case of multiple wells], but the entire functional form, we can proceed as in reference [2] using higher dimensional model representation. However, it is not clear, before doing numerical calculations how meaningful it may be to determine a single unknown [the control laser field] from a very large number of inputs, since numerical approximations in the interaction of the known variables may swamp the individual unknown variable.

Hence the cost functional may be defined in general as

$$C[\vec{a},<\Theta>] = F(<\vec{a},T|\Theta|\vec{a},T> - \Theta_E) + \int_0^T dt [g(<\vec{a},t|\Theta|\vec{a},t> + h(\vec{a}) + \Lambda(\vec{a},t)] \quad \ldots \ldots (11)$$

with F serving as the cost of deviation of the expectation value from the target $\Theta_E$ at the end of period T, g is a function that may do the same throughout the interval [ 0,T], h is a function that accounts for the cost of the parameters, notably the control parameter $a_N$, which can be the laser field, and $\Lambda$ is the function constraining the system to obey Schrodinger equation:

$$\Lambda(t) = <\lambda(t)|[i\hbar\frac{\partial}{\partial t} - H]|\psi(t)> + h.c. \quad \ldots \ldots (12)$$

The optimization of the cost function C leads to a set of equations involving H (and hence **a**), $|\psi(t)>$, $<\Theta>$ and $|\lambda(t)>$. Demanding that $<\Theta>$ remain invariant throughout all *t* with the appropriate choice of the function *g* would again lead to the level set equation (9). The function h(a) which tries to optimize the cost of control would give a set of constraints among the parameters **a** that too would in general be time dependent and hence may be represented by another level set in $\mathbf{R^{n-1}}$ space. The intersection of these two level sets should therefore be the optimal solution. For example, the constraint of minimizing the laser power may be represented by the function

$$h(a_n = E(t)) = (1/2) E(t)^2 \quad \ldots \ldots (13)$$

The general relation for optimal $a_n$ is:

$$\frac{\partial}{\partial <\theta>}[g(<\theta>)]\frac{\partial <\theta>}{\partial a_n} + \frac{\partial h(a)}{\partial a_n} = 0 \quad \ldots \ldots (14)$$

which will lead to a subset of the original level sets (Eq. 9). We have shown such a condition by the blue curve in Fig. 1.

If we do not demand exact constancy of $<\Theta>$ for all time, but periodic return to a preassigned value, then the fixed time function *F* takes over from the continuous time function *g* in Eq. 11. The algorithm now consists of
1) beginning with a guessed $a_n$ and assigned $a_i$ (i =1,2,…n-1)
2) integrating Schrodinger equation to get $|\psi(t)>$,
3) integrating back Lagrange multipliers from t= T, to all t upto t=0.
4) calculating $<\Theta>$ for various ts and the cost functional.
5) changing $a_n$ and finding the gradient $\partial C/\partial a_n$

6) adjusting $a_n$ till the cost is optimal.
7) keeping this value of $a_n$ constant we find values of $\langle\Theta\rangle$ for different **a** on a mesh in $\mathbf{R^{n-1}}$. We make use of HDMR or other interpolation techniques like B-spline to get a level curve/surface.
8) We find the gradient of the surface with respect to each component of **a**, except $a_n$
9) We can thus change the **a** to other level surfaces by changing the system components in time by T and finding the $a_n$ required to give another level surface for the same $\langle\Theta\rangle$.
10) Likewise we also draw the least cost surfaces in $\mathbf{R^{n-1}}$ and the intersection of these surfaces and the constant $\langle\Theta(nT)\rangle$ surfaces gives us the required solution sets.

## IV.    CONCLUSIONS

We have shown that the level set method provides an interesting alternative to the usually complicated inversion process in optimal quantum control. Numerical calculations are in progress with specific potentials and measurables to test the accuracy and reliability of the method in this context.

The author thanks Professor H. Rabitz for helpful discussions.

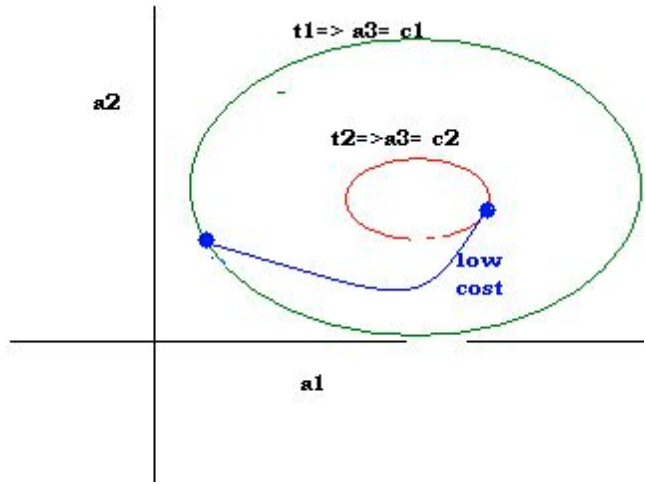

Fig. 1: Two hypothetical level sets when <Θ> depends on two system parameters $a_1$ and $a_2$ and one control parameter $a_3$, which stand in place of time. The blue line is a possible optimal cost trajectory cutting the shown level sets in two points.